\documentclass[fleqn,usenatbib]{mnras}

\usepackage{newtxtext,newtxmath}

\usepackage[T1]{fontenc}
\usepackage{ae,aecompl}

\usepackage{graphicx}    
\usepackage{amsmath}    
\usepackage{enumerate}
\usepackage{hyperref}

\hypersetup{linkcolor=red}          
\title[Explaining the Anisotropy vs Shape Relation]{Physical Explanation for the Galaxy Distribution on the $(\lambda_{\rm R}, \varepsilon)$ and $(V/\sigma, \varepsilon)$ Diagrams or for the Limit on Orbital Anisotropy}

\author[B. Wang et al.]{
	Bitao Wang,$^{1,2,3}$\thanks{bt-wang@pku.edu.cn}
	Michele Cappellari,$^{3}$
	Yingjie Peng$^{2}$\thanks{yjpeng@pku.edu.cn}
	\\
	$^{1}$Department of Astronomy, School of Physics, Peking University, Beijing 100871, China\\
	$^{2}$Kavli Institute for Astronomy and Astrophysics, Peking University, Beijing 100871, China\\
	$^{3}$Sub-department of Astrophysics, Department of Physics, University of Oxford, Denys Wilkinson Building, Keble Road, Oxford OX1 3RH, UK
}

\date{}

\pubyear{2020}

\begin{document}
\label{firstpage}
\pagerange{\pageref{firstpage}--\pageref{lastpage}}
\maketitle

\begin{abstract}
In the $(\lambda_{\rm R}, \varepsilon)$ and $(V/\sigma, \varepsilon)$ diagrams for characterizing dynamical states, the fast-rotator galaxies (both early-type and spirals) are distributed within a well-defined leaf-shaped envelope.
This was explained as due to an upper limit to the orbital anisotropy increasing with galaxy intrinsic flattening.
However, a physical explanation for this empirical trend was missing.
Here we construct Jeans Anisotropic Models (JAM), with either cylindrically or spherically aligned velocity ellipsoid (two extreme assumptions), and each with either spatially-constant or -variable anisotropy.
We use JAM to build mock samples of axisymmetric galaxies, assuming on average an oblate shape for the velocity ellipsoid (as required to reproduce the rotation of real galaxies), and limiting the radial anisotropy $\beta$ to the range allowed by physical solutions.
We find that all four mock samples naturally {\em predict} the observed galaxy distribution on the $(\lambda_{\rm R}, \varepsilon)$ and $(V/\sigma, \varepsilon)$ diagrams, without further assumptions. Given the similarity of the results from quite different models, we conclude that the empirical anisotropy upper limit in real galaxies, and the corresponding observed distributions in the $(\lambda_{\rm R}, \varepsilon)$ and $(V/\sigma, \varepsilon)$ diagrams, are due to the lack of physical axisymmetric equilibrium solutions at high $\beta$ anisotropy when the velocity ellipsoid is close to oblate.
\end{abstract}



\section{Introduction}\label{sec:intro}

Gravitation dominates in galaxies and makes thermal equilibrium unattainable, so that their current configurations cannot be simply explained as states of maximum entropy \citep[e.g.][]{2008gady.book.....B}.
To understand the present states of galaxies it requires the knowledge about the initial conditions of their formation and the subsequent dynamical processes they experienced.
Important clues to the assembly histories can be held in stellar kinematics \citep[e.g. review by][]{2016ARA&A..54..597C}.

One way to characterize the stellar kinematics is the anisotropy of the orbital distribution quantified through the ratio of orthogonal velocity dispersions.
Decades ago, elliptical galaxies were thought to be isotropic and flattened by rotation \citep[e.g.,][]{1975ApJ...201..296G}.
However, long-slit spectra of bright elliptical galaxies revealed too low rotation velocities against their presumed isotropy \citep{1975ApJ...200..439B, 1977ApJ...218L..43I, 1978MNRAS.183..501B}.
Orbital anisotropy was proposed to explain the low levels of rotation observed at different galaxy shapes \citep{1976MNRAS.177...19B, 1978MNRAS.183..501B}.
Following works were extended onto small samples of fainter ellipticals \citep{1983ApJ...266...41D} and bulges of spiral galaxies \citep{1982ApJ...256..460K, 1982ApJ...257...75K} which showed that most of them had rotation comparable to isotropic rotators.
While more recent works also indicated the existence of fainter early-type galaxies (ETGs) with significant anisotropy \citep{2007MNRAS.379..418C, 2009MNRAS.393..641T}.

Using three-integral axisymmetric toy models, \citet{2009MNRAS.393..641T} shows that at fixed flattening galaxies achieve higher entropy with larger velocity anisotropy.
Therefore, perturbations which move galaxies away from equilibrium may lead to configurations of larger anisotropy.
Structures on the disks such as giant molecular clouds, bars and spiral arms can perturb stars and heat them anisotropically \citep{1951ApJ...114..385S, 1990MNRAS.245..305J, 2003AJ....126.2707S}.
Mergers may enhance the vertical velocity dispersion relative to the dispersions on equatorial plane \citep{1992ApJ...389....5T, 2004MNRAS.351.1215B}.
\citet{2019MNRAS.485..972T} reports that the anisotropy also correlates with the intrinsic shapes of inner dark matter halos.

Based on tensor virial theorem, for a given intrinsic ellipticity galaxies can display any bulk rotation velocity between a maximum value and no rotation at all \citep{2005MNRAS.363..937B}.
Lower rotations can be achieved by increasing the anisotropy up to a maximum theoretical value at $V/\sigma=0$.
However, the first statistically-significant set of three-integral axisymmetric \citet{Schwarzschild1979} models of galaxies based on integral-field stellar kinematics revealed that real galaxies do not reach the maximum anisotropy allowed by the tensor virial theorem, but instead lie below a limit $\beta\la 0.7\times\varepsilon_{\rm intr}$ \citep[the `magenta line' of][]{2007MNRAS.379..418C}.
This upper limit appeared consistent with the observed lower boundary of the distribution of fast rotators on the $(V/\sigma, \varepsilon)$ diagram.

The distribution of galaxy samples with ever-increasing size \citep{2011MNRAS.414..888E, 2018MNRAS.477.4711G, 2020MNRAS.495.1958W} on either the $(V/\sigma, \varepsilon)$ diagram by \citet{2005MNRAS.363..937B} or the $(\lambda_{\rm R},\varepsilon)$ diagram by \citet{2007MNRAS.379..401E}, unambiguously confirmed that galaxies follow the leaf-like distribution predicted by randomly-oriented axisymmetric models with anisotropy upper limit that increases with their intrinsic flattening \citep[sec.~3 in the review by][]{2016ARA&A..54..597C}.
However, a physical explanation for the existence of this empirical relation between anisotropy and shape was not known and we try to find it in this Letter by exploiting Jeans Anisotropic Models (JAM) of galaxy dynamics.


\section{Data and Models}\label{sec:data}

The dynamic models used in this work are solutions for steady-state axisymmetric Jeans equations of velocity second moments \citep{1922MNRAS..82..122J}, allowing for velocity anisotropy.
Specifically, the equation solutions\footnote{We used v6.2 of the \textsc{jampy} Python software package available from \url{https://pypi.org/project/jampy/}} based on the Multi-Gaussian Expansion \citep[MGE; ][]{1994A&A...285..723E, 2002MNRAS.333..400C} formalism are given in \citet{2008MNRAS.390...71C} and \citet{Cappellari2020} making the two extreme assumptions of a cylindrically-aligned (JAM$_{\rm cyl}$) and a spherically-aligned (JAM$_{\rm sph}$) velocity ellipsoid respectively.
The two models with different velocity ellipsoid alignment are characterized by anisotropy parameters of orthogonal velocity dispersions \citep{1982MNRAS.200..361B, 2008gady.book.....B}:
\begin{align}
\beta_\mathrm{cyl}(R,z)& \equiv \, 1-(\sigma_z/\sigma_R)^2&
\beta_\mathrm{sph}(r,\theta)& \equiv \, 1-(\sigma_{\theta}/\sigma_r)^2\\
\gamma_\mathrm{cyl}(R,z)& \equiv \, 1-(\sigma_{\phi}/\sigma_R)^2&
\gamma_\mathrm{sph}(r,\theta)& \equiv \, 1-(\sigma_{\phi}/\sigma_r)^2
\end{align}

We assume spatially constant total mass-to-light ratio (M/L), which is approximate.
But even accounting for stellar-M/L gradients and the dark matter, within one half-light radius ($R_{\rm e}$) where our measurements will be carried out, the profiles of total mass indeed closely follow those of the stellar mass \citep[e.g. fig.~10 of][]{Poci2017}.
This explains the success of the mass-follow-light models in describing the integral-field kinematics of real galaxies \citep{2013MNRAS.432.1709C}.
The contribution of central supermassive black holes is ignored due to its minimal influence on the kinematics of the scale that we are interested in.

We build dynamic models based on realistic galaxy light distributions.
The MGE photometric models of r-band light of 112 regular-rotator $\mathrm{ATLAS}^\mathrm{3D}$ ETGs are taken from \citet{2013MNRAS.432.1894S}.
This is a subsample of the 260 $\mathrm{ATLAS}^\mathrm{3D}$ ETGs and includes all fast rotators (as flagged `F' in table~B1 by \citealt{2011MNRAS.414..888E}) with high inclination ($i>60^\circ$).
The inclination was measured via JAM dynamic modelling and taken from table~1 of \citealt{2013MNRAS.432.1709C}, including only reliable measurements (`quality' $>0$).
This inclination criterion is meant to reduce the uncertainty in the mass-deprojection degeneracy \citep{Rybicki1987}, crucial for deriving the intrinsic density profiles of galaxies.
The focus of this study is the leaf-like envelope of the fast rotators.
So we exclude slow rotators as they are generally triaxial \citep{2016ARA&A..54..597C} and do not follow the distribution of the fast rotators.
And reproducing the envelope requires adequate coverage of intrinsic ellipticities of the real galaxies that form the envelope, which is satisfied by our sample.

\begin{figure}
	\includegraphics[width=0.49\textwidth,trim={2.7cm 2.3cm 3.4cm 2.5cm},clip]{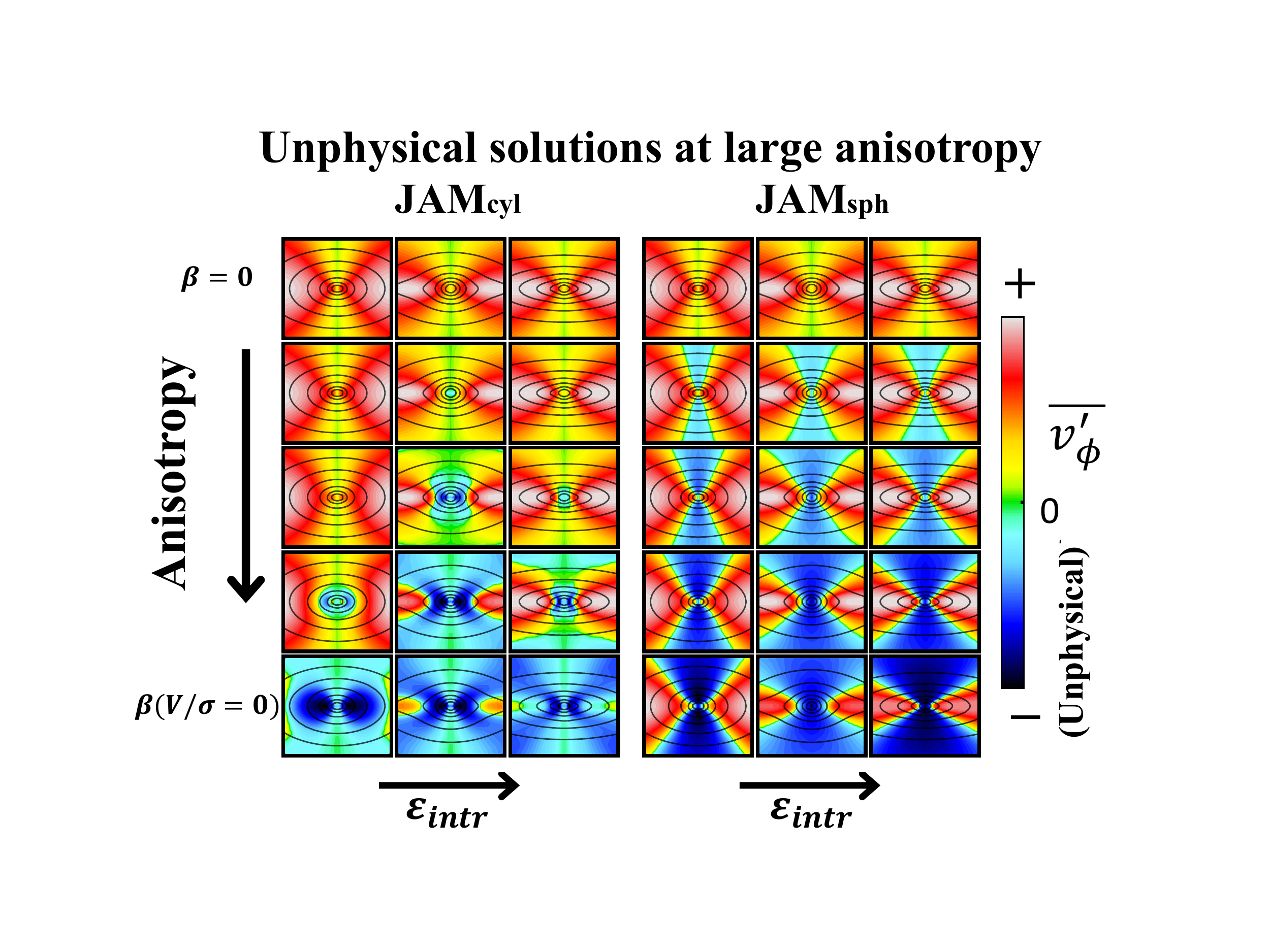}
	\caption{$\overline{v'_{\phi}}\,(R,z)=\mathrm{sign}(\overline{v_{\phi}}^2)\times|\overline{v_{\phi}}^2|^{1/2}$ maps derived with $\mathrm{JAM}_\mathrm{cyl}$ (left panels) and $\mathrm{JAM}_\mathrm{sph}$ (right panels) for three galaxies with $\varepsilon_\mathrm{intr}$ increasing from left to right (NGC 4551, NGC 4474 and NGC 0448).
	$\beta$ increases from zero to the maximum value allowed by tensor virial theorem for certain $\varepsilon_\mathrm{intr}$.
	In each panel, overlaid black lines show the density contours.
	}
	\label{fig:LB}
\end{figure}


For a direct comparison with observation, we measure ellipticity, $(V/\sigma)_{\rm e}$ and $\lambda_{\rm R_e}$ as is done for real galaxies using integral-field data.
The effective ellipticity $\varepsilon$ and $(V/\sigma)_{\rm e}$ are measured within the half-light isophote as defined in eq.~(10)--(11) of \citet{2007MNRAS.379..418C}. This ellipticity is measured from the MGE parametrization of the surface brightness with the routine \textsc{mge\_half\_light\_isophote}\footnote{Also included in the \textsc{jampy} Python software package.} which implements the steps (i)-(iv) above equation (12) in \citet{2013MNRAS.432.1709C}.
The specific angular momentum proxy $\lambda_{\rm R_e}$ is computed as defined in eq.~(1)--(2) of \citet{2007MNRAS.379..401E}.

\section{Non-physical Models at Large Anisotropy}\label{sec:effect}

\citet{2007MNRAS.379..418C} found that real galaxies have $\gamma_\mathrm{cyl}$ around zero and the main trend of anisotropy with flattening is driven by the systematic change of $\beta_\mathrm{cyl}$.
This on-average oblate shape (i.e. $\sigma_{\phi} \sim \sigma_{R}$) of the velocity ellipsoid in fast-rotator ETGs is strikingly apparent in fig.~11 of \citet{2016ARA&A..54..597C}.

If one fixes $\gamma$ and increases $\beta$, the models become unphysical when the {\em squared} streaming velocities $\overline{v_{\phi}}^2=\overline{v_{\phi}^2}-\sigma_{\phi}^2$ is significantly negative in non-negligible parts of the models.
Examples are given in \autoref{fig:LB} which shows maps\footnote{For $\overline{v_{\phi}}^2>0$, the $\overline{v'_{\phi}}$ is the usual streaming velocity, while for $\overline{v_{\phi}}^2<0$, the $\overline{v'_{\phi}}$ is the absolute value of the complex $\overline{v_{\phi}}$, but we give it a negative sign to indicate it is unphysical.} of $\overline{v'_{\phi}}\,(R,z)\equiv\mathrm{sign}(\overline{v_{\phi}}^2)\times|\overline{v_{\phi}}^2|^{1/2}$ in a $1R_e\times 1R_e$ region and in a $(R,z,\phi)$ cylindrical coordinate system where the $z$ axis is the galaxy symmetry axis.
For each galaxy, $\beta$ increases with a certain linear step from $\beta=0$ at the top panel to the maximum anisotropy allowed by the tensor virial theorem $\beta(V/\sigma=0)$ at the bottom.
The colour bar range of $\overline{v'_{\phi}}$ is symmetric about zero so that unphysical areas have blue colours.
Note that unphysical models are expected as the Jeans equations themselves do not guarantee physically meaningful solutions.

Isotropic models (the first row) are entirely physical with non-negative values of $\overline{v'_{\phi}}$ everywhere.
At certain large values of $\beta$, parts of the models start having significantly negative (dark blue) $\overline{v'_{\phi}}$.
These unphysical regions grow with further increased $\beta$.
The anisotropy at which a model becomes mildly unphysical can be considered as a natural upper limit for $\beta$.

\section{Predicted $(V/\sigma, \varepsilon)$ and $(\lambda_{\rm R},\varepsilon)$ distributions}\label{sec:mc}

We carry out Monte Carlo simulations to model the distribution of galaxies on the $(\beta,\varepsilon_\mathrm{intr})$, $(\lambda_{\rm R},\varepsilon)$ and $(V/\sigma,\varepsilon)$ diagrams similarly to what was done in appendix~C of \citet{2007MNRAS.379..418C} or appendix~B of \citet{2011MNRAS.414..888E}. The key difference, and the novelty of this paper, is that in our case the anisotropy $\beta$ of each galaxy is not assumed but comes directly from the requirement of a physical JAM solution for each galaxy.

\subsection{Modelling the tangential anisotropy}\label{sec:kappa}

A crucial aspect of our simulations is the choice for the distribution of tangential anisotropy $\gamma$. It is clear that one can construct physical axisymmetric galaxy models with arbitrarily low level of $V/\sigma$ or $\lambda_{\rm R}$ by allowing for counter-rotating disks \citep[sec.~3.4.3 of][]{2016ARA&A..54..597C} which have large $\gamma$ and fall well below the leaf-like envelope populated by fast rotators. However, observationally counter-rotating disks are rare and the area below the leaf-like envelope in the $(V/\sigma,\varepsilon)$ and $(\lambda_{\rm R},\varepsilon)$ diagrams is sparsely populated.

To make sure that our mock galaxies match the $\gamma$ anisotropy of real galaxies, we require them to reproduce both (i) the measured range of $\gamma$ for fast rotators, from Schwarzschild models, in fig.~2 of \citet{2007MNRAS.379..418C} and (ii) the distribution of rotation parameter $\kappa$, from Jeans models, in fig.~11 of \citet{2016ARA&A..54..597C}. Here $\kappa$ is defined by equation (52) in \citet{2008MNRAS.390...71C} as the ratio of observed rotation and the rotation of a model with oblate velocity ellipsoid, and can be considered as a quantification of the tangential anisotropy.

\begin{figure}
	\includegraphics[width=0.49\textwidth]{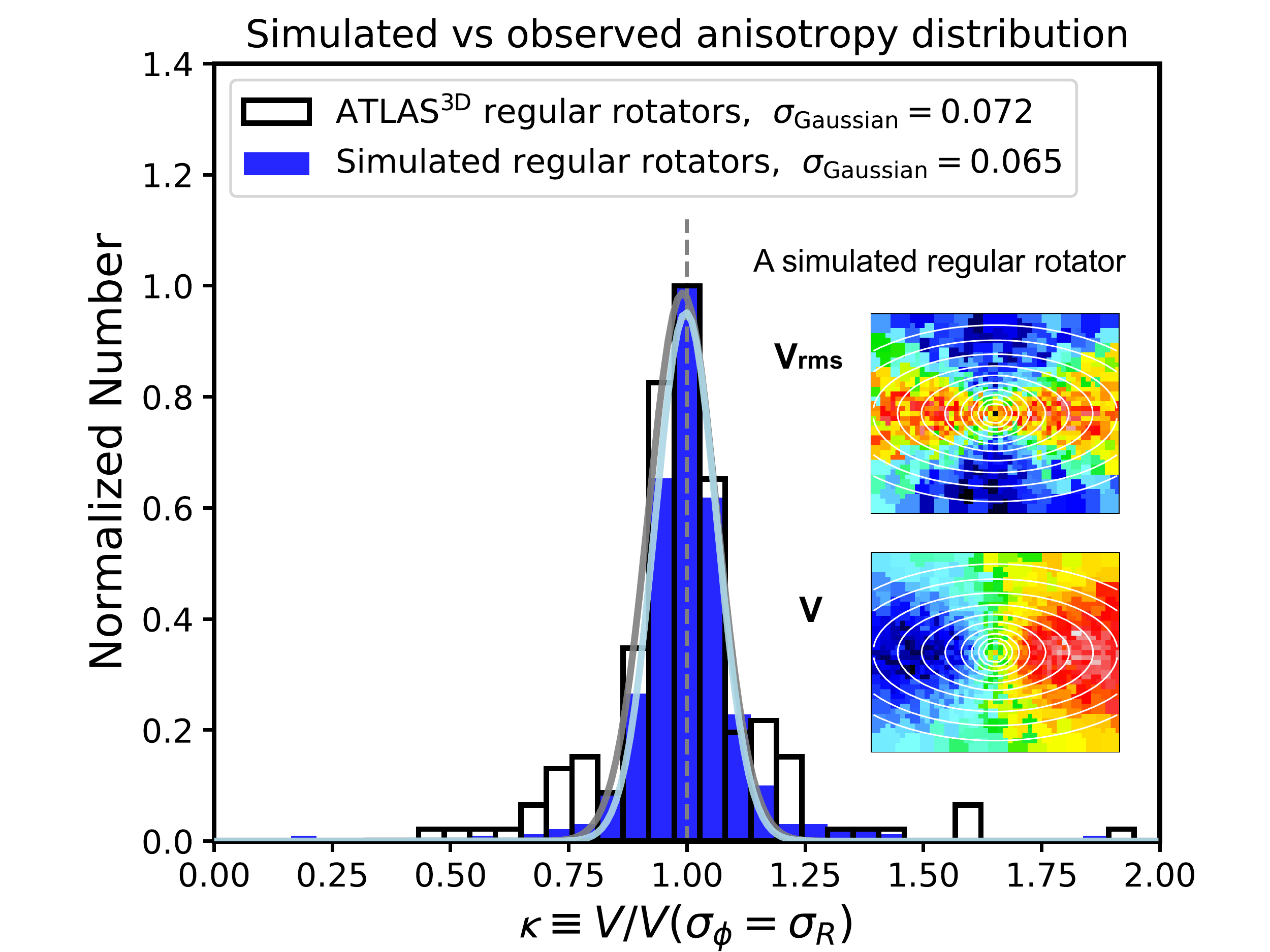}
	\caption{The consistency in the anisotropy $\kappa$ distributions of simulated regular rotators (blue) under our assumed $\gamma$ distribution and the $\mathrm{ATLAS}^\mathrm{3D}$ regular rotators (white).
			The histograms are normalized by the peaks and the light blue and gray curves are the best fitting Gaussians for the simulation and observation respectively.
			The velocity and root-mean-square velocity of a projected regular rotator are shown as an example of the mock data.
	}
	\label{fig:kappa}
\end{figure}

Closely mimicking the observation, in our simulations each projected model (described in \autoref{sec:con}) is spatially Voronoi binned \citep{2003MNRAS.342..345C} and Gaussian noises with dispersions $\varepsilon_V = 0.1 V_\mathrm{max}$ and $\varepsilon_{\sigma}= 0.1 \sigma$ are added to the kinematics, producing realistic maps (see an example in \autoref{fig:kappa}). The JAM$_\mathrm{cyl}$ models are then treated as mock observations by first fitting the $V_{\rm rms}$ for $\beta$, inclination and M/L and then fitting the V for $\kappa$.

We found that we can reproduce the above two anisotropy observations by adopting a Gaussian distribution for the ratio $\sigma_{\phi}/\sigma_{R}$ with mean $\mu=1$ and dispersion $\sigma=0.07$.
This results in a $\gamma$ distribution with tails extending to $\gamma\approx\pm 0.2$, consistent with \citet{2007MNRAS.379..418C}, and a distribution of $\kappa$ that quantitatively matches the observations (\autoref{fig:kappa}).


\subsection{Models with spatially-constant anisotropy}\label{sec:con}

\begin{figure*}
	\includegraphics[width=0.49\textwidth]{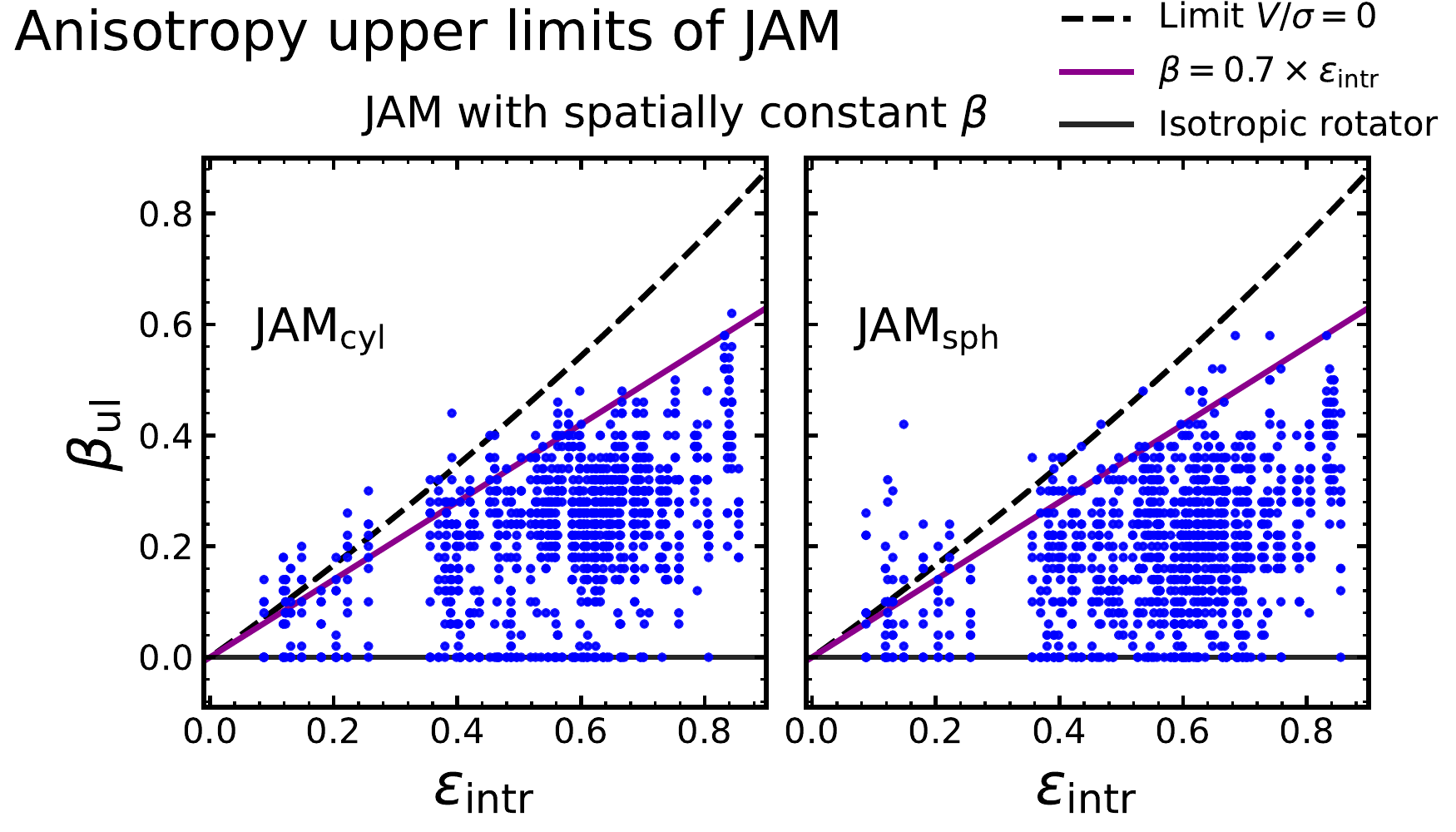}
	\includegraphics[width=0.49\textwidth]{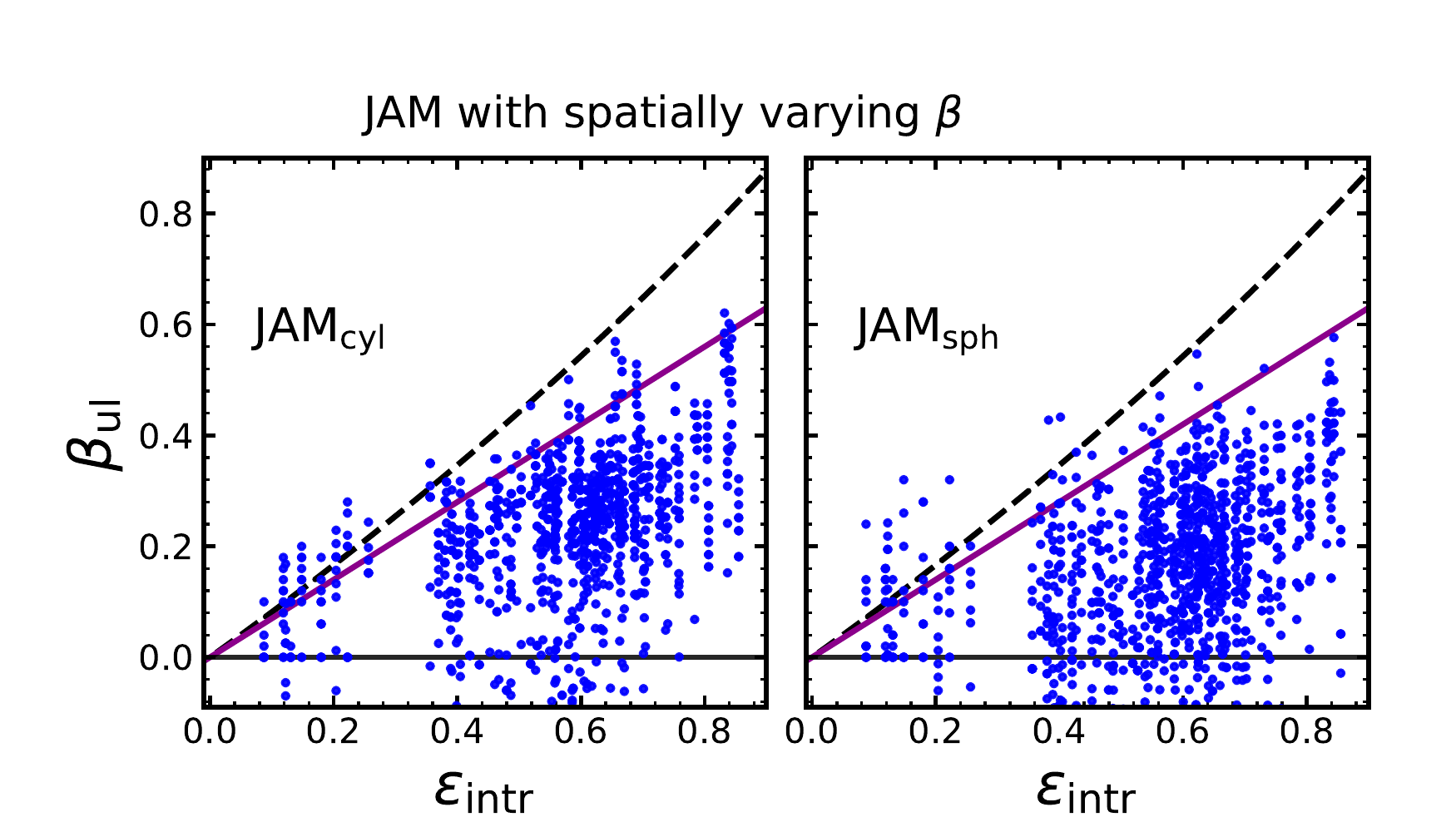}
	\caption{Anisotropy upper limit $\beta_\mathrm{ul}$ as a function of intrinsic ellipticity $\varepsilon_\mathrm{intr}$ of JAM models assuming spatially constant $\beta$ (left panels) and spatially varying $\beta$ (right panels).
		In each case, cylindrically and spherically aligned velocity ellipsoids are assumed respectively.
		$\beta_\mathrm{ul}$ is the anisotropy from where the model (based on the density profile of a certain galaxy) is considered mildly unphysical and thus unrepresentative for real galaxies.
		For the models with spatially varying $\beta$, $\beta_\mathrm{ul}$ is integrated within the half-light ellipse.
		The upper bound of $\beta_\mathrm{ul}$ in each panel roughly matches the magenta line, an empirical upper limit from \citet{2007MNRAS.379..418C}, and is below the maximum allowed by tensor virial theorem under $\gamma=0$ (the black dashed line).
	}
	\label{fig:be}
\end{figure*}

\begin{figure*}
	\includegraphics[width=0.49\textwidth]{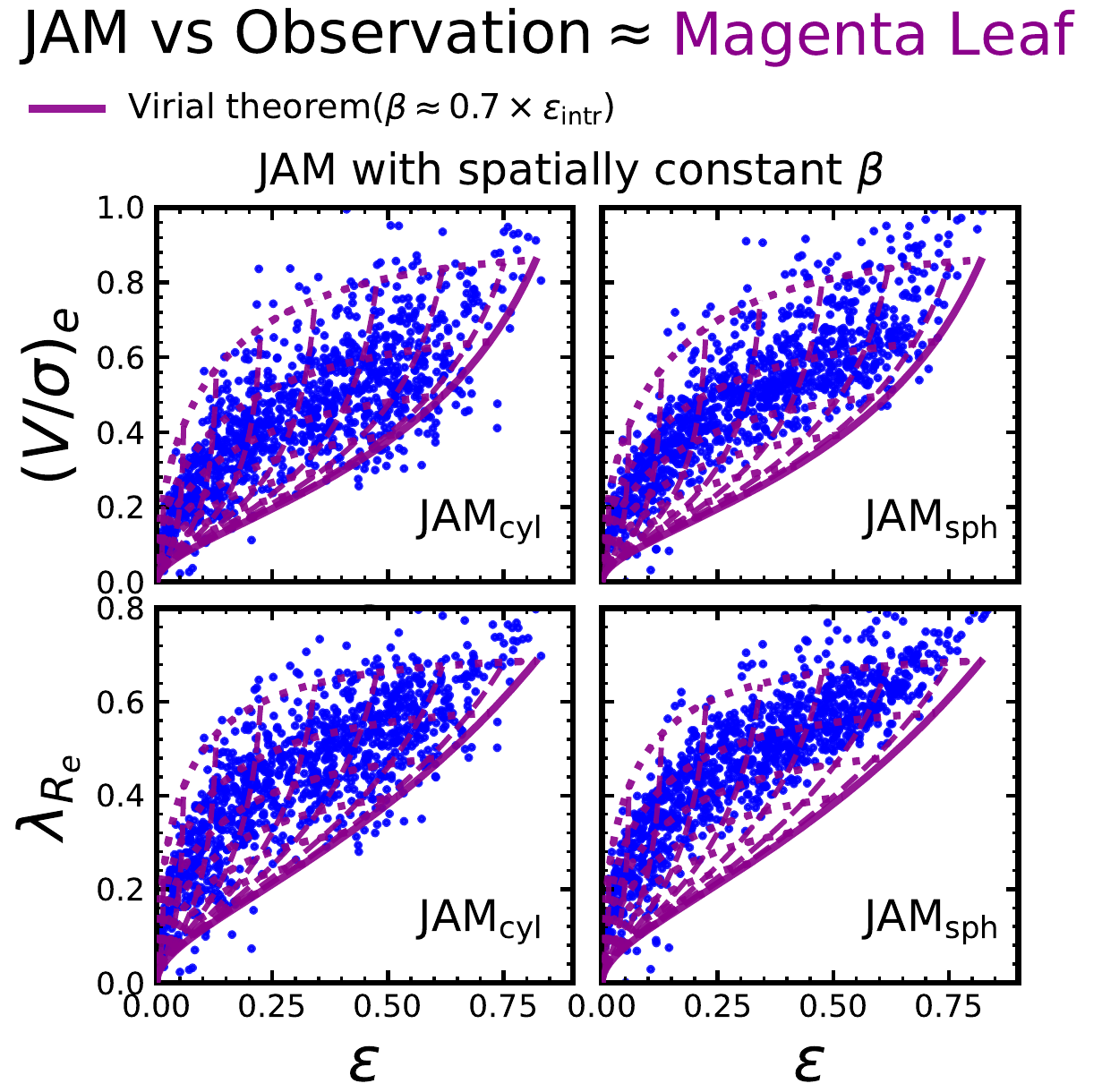}
	\includegraphics[width=0.49\textwidth]{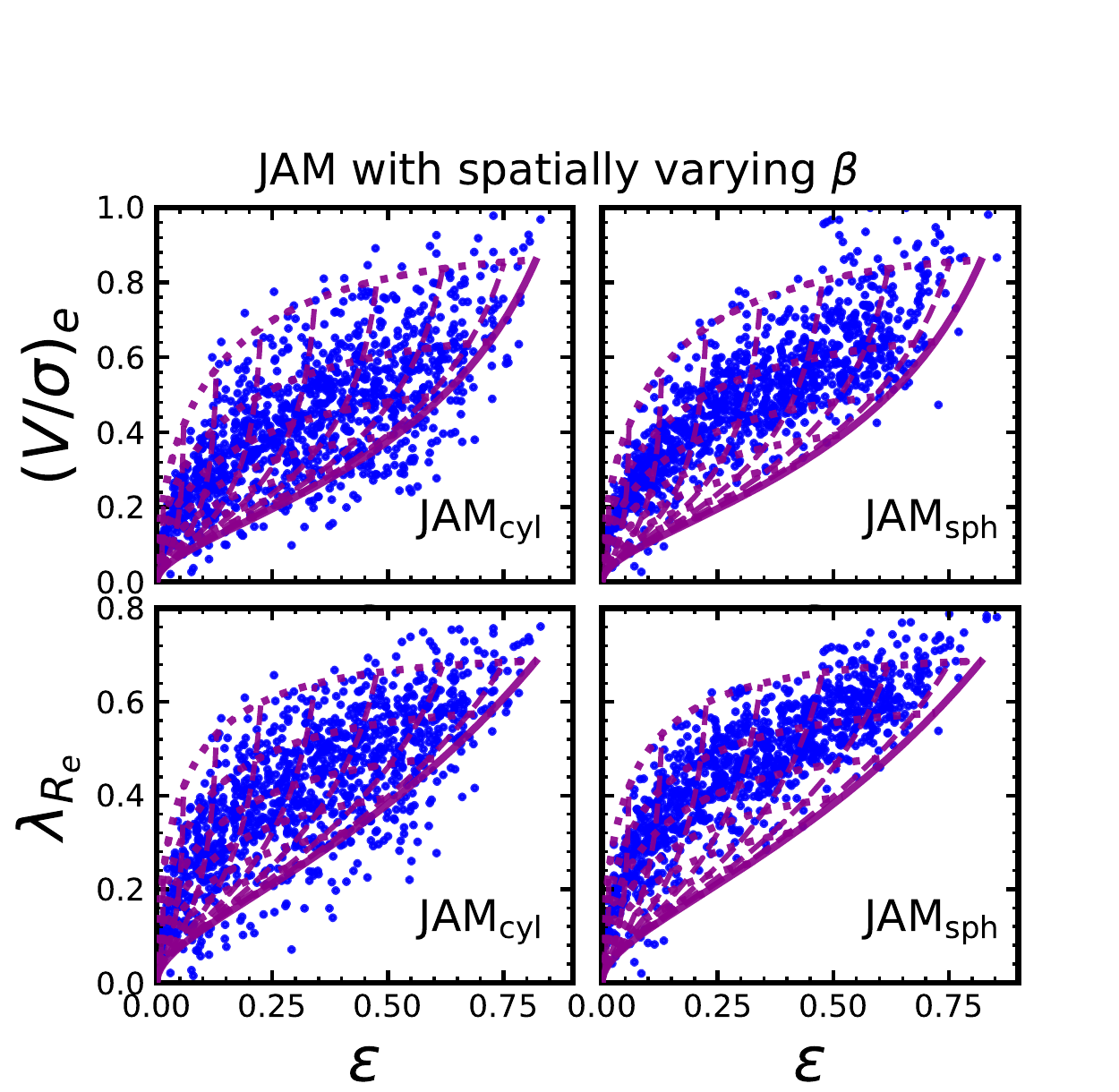}
	\caption{$(V/\sigma)_{\rm e}$ (upper row) and $\lambda_{\rm R_e}$ (lower row) as a function of ellipticity $\varepsilon$ for JAM models projected at random inclinations, assuming spatially constant $\beta$ (left panels) and spatially varying $\beta$ (right panels).
	Again, in each case cylindrically and spherically aligned velocity ellipsoids are assumed respectively.
	Anisotropy $\beta$ of each model is restricted by the physical upper limit $\beta_\mathrm{ul}$ which is determined respectively for each model.
	Virial theorem predictions for oblate rotators following the anisotropy relation $\beta \approx \delta = 0.7 \times \varepsilon _{\mathrm{intr}}$ of \citet{2007MNRAS.379..418C} are shown, with the solid magenta line showing an edge-on view at different intrinsic ellipticities, and the dotted and dashed magenta lines for other inclinations.
	Note that this "magenta leaf" envelope approximately covers the area populated by real galaxies in this $\varepsilon_\mathrm{intr}$ range.
	}
	\label{fig:reprod}
\end{figure*}

For each galaxy in our sample, we compute 10 models based on its deprojected MGE density distribution.
This is done separately for both the $\mathrm{JAM}_\mathrm{cyl}$ and the $\mathrm{JAM}_\mathrm{sph}$ models with the intrinsic kinematics computed with the procedure \textsc{jam\_axi\_intr}, using the keyword \texttt{align=`cyl'} and \texttt{align=`sph'} respectively.

For every model, we start by drawing a value of the ratio $\sigma_{\phi}/\sigma_{R}$ for JAM$_\mathrm{cyl}$ (or $\sigma_{\phi}/\sigma_{r}$ for JAM$_\mathrm{sph}$) from the Gaussian distribution determined in \autoref{sec:kappa}.
With this sampled $\gamma$, then we try a sequence of $\beta$ values starting from 0 and increasing with a step 0.02, to find out at which value the model meets our `mildly unphysical' criterion.
We have tried a variety of slightly different `mildly unphysical' criteria: (i) $|\mathrm{min}(\overline{v'_{\phi}})|>f_1\times\mathrm{max}(\overline{v'_{\phi}})$ (the peak unphysical velocity is no longer  small compared with the physical one); (ii) $f_{\,V_\mathrm{neg}}>f_2$ (the volume fraction of unphysical part of the model is no longer  small); (iii) the two criteria combined, where $f_1$ and $f_2$ are constants; or (iv) the fraction of volume where $|\mathrm{min}(\overline{v'_{\phi}})|>f_1\times\mathrm{max}(\overline{v'_{\phi}})$ is larger than $f_2$.
And for direct comparison with observation, we only take into account the part of the model enclosed within the half-light ellipse in the $(R,z)$ plane.
We obtained qualitatively similar results with all these different criteria, but in the following, we adopted the first one (i) as our standard criterion with $f_1=0.2$.
This criterion typically corresponds to an unphysical volume fraction of several per cent inside the half-light ellipse, which indeed indicates an at most mildly unphysical model.

Given that nearly every physical model is only a simplified version of reality, it does not make sense to define the model as unphysical as soon as the first values of $\overline{v_{\phi}}^2$ become negative. This would lead to an unrealistically too strict criterion, as we try to qualitatively approximate what may happen in real galaxies and are not interested in the mathematical aspects of the JAM solutions.

After finding the upper limit $\beta_\mathrm{ul}$, for the model we uniformly sample a value for $\beta$ in the range $[0,\beta_\mathrm{ul}]$.
This assumes that real galaxies can possess the full range of anisotropy allowed by physical solutions.
Biasing the sampling toward low or high anisotropy gives qualitatively the same results.
Lastly, we draw a random orientation on the sphere of viewing angles. We use the JAM procedure \textsc{jam\_axi\_proj} to compute the predicted kinematics projected along the line-of-sight, out of which we measure $(V/\sigma)_{\rm e}$ and $\lambda_{\rm R_e}$ using the standard approach as used in observation.
Note that, during the line-of-sight integration, the \textsc{jam\_axi\_proj} procedure sets $\overline{v_{\phi}}=0$ to the unphysical part. This makes little difference to the projected kinematics, given that we only consider mildly unphysical models and the volume fraction of the unphysical part is typically only several percent inside $R_e$.

The left two panels of \autoref{fig:be} show the final distribution of $\beta_\mathrm{ul}$ as a function of $\varepsilon_\mathrm{intr}$ for $\mathrm{JAM}_\mathrm{cyl}$ and $\mathrm{JAM}_\mathrm{sph}$ respectively.
In the panels, the magenta line ($\beta=0.7\varepsilon_\mathrm{intr}$) is the empirical upper limit based on both the Schwarzschild models of \citet{2007MNRAS.379..418C} and the Jeans models of \citet{2013MNRAS.432.1709C}.
While the black dashed line is the zero rotation limit set by tensor virial theorem when $\gamma = 0$.

The results indicate that the magenta line approximately corresponds to the upper limit for the JAM models under the condition of being physical.
The tolerance of velocity anisotropy varies significantly between different galaxy density profiles and only some of them have $\beta_\mathrm{ul}$ close to the magenta line.

The distributions of projected models on $(V/\sigma,\varepsilon)$ and $(\lambda_{\rm R},\varepsilon)$ planes are shown in the left four panels of \autoref{fig:reprod}.
Under the conditions $\beta = 0.7 \varepsilon _{\mathrm{intr}}$ and $\gamma=0$, the edge-on prediction (the solid magenta line in \autoref{fig:reprod}) from tensor virial theorem together with its projections at different inclinations (magenta dashed and dotted lines) form the envelope (the `magenta leaf' of \citealt{2007MNRAS.379..418C}) that matches well the observed distribution boundaries of $\mathrm{ATLAS}^\mathrm{3D}$ \citep{2007MNRAS.379..401E}, MaNGA \citep{2018MNRAS.477.4711G} and SAMI \citep{vandeSande2017} galaxies (see more about the theoretical tracks in section 3.5 of \citealt{2016ARA&A..54..597C}).

The resultant distributions of $\mathrm{JAM}_\mathrm{cyl}$ models in the first column highly resemble the ones of real galaxies which are represented by the magenta envelopes.
A similarity between observations and models is also visible in the second column for $\mathrm{JAM}_\mathrm{sph}$ models, but to a lesser degree.
This may imply that the velocity ellipsoid of real galaxies is on average better described by cylindrically- than spherically-aligned models.

\subsection{Models with spatially-variable anisotropy}\label{sec:var}

While a spatially constant anisotropy $\beta$ is assumed previously, in real galaxies $\beta$ can vary with spatial position.
However there are no systematic analyses of the  anisotropy variation in fast rotators. Studies of a handful of galaxies with high-quality integral-field stellar kinematics have found that, beyond the sphere of influence of the central supermassive black hole, the $\sigma_z/\sigma_R$ ratio varies on the order of 20\% within $R_{\rm e}$ \citep[e.g.,][]{2008IAUS..245..215C, 2018MNRAS.477.3030K}.

To model the $\beta$ variation, we assume that the rounder bulges and stellar halos are more isotropic than the discs.
And in the previous Monte Carlo simulation, while searching for $\beta_\mathrm{ul}$ we multiply the dispersion ratio (e.g. $\sigma_z/\sigma_R$ for JAM$_\mathrm{cyl}$) by a factor of 1.1 for the inner or rounder MGE components (the Gaussians with $\sigma<0.3R_{\rm e}$ or axial ratio $q>0.7$) and a factor of 1/1.1 for the remaining flatter and outer components.
The typical change of $\beta$ is then $\sim 0.2$ inside one $R_e$, increasing outward.
When the mildly unphysical criterion is met, we record the flux weighted $\beta$ inside one $R_e$ as $\beta_\mathrm{ul}$.
The results under spatially varying $\beta$ are shown in the right panels of \autoref{fig:be} and \autoref{fig:reprod}.
No significant difference is seen compared with the results under spatially constant $\beta$.

\section{Conclusions}\label{sec:discuss}

In this letter, we have used Jeans anisotropic models (JAM) in combination with realistic stellar density distributions of $\mathrm{ATLAS}^\mathrm{3D}$ galaxies to try to understand the physical origin for the observed distribution of fast-rotator galaxies (spirals and ETGs) on both the $(\lambda_{\rm R_e},\varepsilon)$ and the $((V/\sigma)_e,\varepsilon)$ diagrams, and for the empirical upper limit on the radial anisotropy $\beta$ as a function of the galaxy intrinsic flattening.

We found that if we adopt an on-average oblate velocity ellipsoid, as constrained by the observations, and require our models to be no more than mildly unphysical (i.e. at most having only weakly negative $\overline{v_{\phi}}^2$), and randomly project the models on the plane of the sky, we can naturally reproduce the observed distributions of galaxies without the need to make additional assumptions about the galaxy anisotropy. This is true for two extreme assumptions on the orientation of the velocity ellipsoid (either cylindrically or spherically aligned) and for both spatially-constant and -variable anisotropy. The result remains qualitatively similar for different criteria to define an unphysical model.

Although our models only approximately describe real galaxies, the robustness of the qualitative result against the different assumptions suggests that the same general phenomenon may apply to real galaxies.
We conclude that the leaf-like distribution of galaxies on the $(\lambda_{\rm R_e},\varepsilon)$ and the $((V/\sigma)_e,\varepsilon)$ diagrams, as well as the empirical upper limit on the radial anisotropy $\beta$, are due to the lack of physical equilibrium solutions at large $\beta$ among regular rotators. The only way that fast-rotator galaxies appear to reach the lowest levels of rotation is when the galaxies contain counterrotating disks, which are rare in the general population.

\vspace{-0.5cm}
\section*{Acknowledgements}
We thank our referee for the thoughtful comments.
BW acknowledges the financial support from the China Scholarship Council during his stay in Oxford.
YP acknowledges the National Key R\&D Program of China, Grant 2016YFA0400702 and NSFC Grant No. 11773001, 11721303, 11991052.
\vspace{-0.5cm}

\section*{Data Availability}
The MGE photometric models used in this work are available from \url{https://purl.org/atlas3d}

\vspace{-0.3cm}



\bibliographystyle{mnras}

\begin{thebibliography}{}
\makeatletter
\relax
\def\mn@urlcharsother{\let\do\@makeother \do\$\do\&\do\#\do\^\do\_\do\%\do\~}
\def\mn@doi{\begingroup\mn@urlcharsother \@ifnextchar [ {\mn@doi@}
  {\mn@doi@[]}}
\def\mn@doi@[#1]#2{\def\@tempa{#1}\ifx\@tempa\@empty \href
  {http://dx.doi.org/#2} {doi:#2}\else \href {http://dx.doi.org/#2} {#1}\fi
  \endgroup}
\def\mn@eprint#1#2{\mn@eprint@#1:#2::\@nil}
\def\mn@eprint@arXiv#1{\href {http://arxiv.org/abs/#1} {{\tt arXiv:#1}}}
\def\mn@eprint@dblp#1{\href {http://dblp.uni-trier.de/rec/bibtex/#1.xml}
  {dblp:#1}}
\def\mn@eprint@#1:#2:#3:#4\@nil{\def\@tempa {#1}\def\@tempb {#2}\def\@tempc
  {#3}\ifx \@tempc \@empty \let \@tempc \@tempb \let \@tempb \@tempa \fi \ifx
  \@tempb \@empty \def\@tempb {arXiv}\fi \@ifundefined
  {mn@eprint@\@tempb}{\@tempb:\@tempc}{\expandafter \expandafter \csname
  mn@eprint@\@tempb\endcsname \expandafter{\@tempc}}}

\bibitem[\protect\citeauthoryear{{Benson}, {Lacey}, {Frenk}, {Baugh}  \&
  {Cole}}{{Benson} et~al.}{2004}]{2004MNRAS.351.1215B}
{Benson} A.~J.,  {Lacey} C.~G.,  {Frenk} C.~S.,  {Baugh} C.~M.,   {Cole} S.,
  2004, \mn@doi [\mnras] {10.1111/j.1365-2966.2004.07870.x}, \href
  {https://ui.adsabs.harvard.edu/abs/2004MNRAS.351.1215B} {351, 1215}

\bibitem[\protect\citeauthoryear{{Bertola} \& {Capaccioli}}{{Bertola} \&
  {Capaccioli}}{1975}]{1975ApJ...200..439B}
{Bertola} F.,  {Capaccioli} M.,  1975, \mn@doi [\apj] {10.1086/153808}, \href
  {https://ui.adsabs.harvard.edu/abs/1975ApJ...200..439B} {200, 439}

\bibitem[\protect\citeauthoryear{{Binney}}{{Binney}}{1976}]{1976MNRAS.177...19B}
{Binney} J.,  1976, \mn@doi [\mnras] {10.1093/mnras/177.1.19}, \href
  {https://ui.adsabs.harvard.edu/abs/1976MNRAS.177...19B} {177, 19}

\bibitem[\protect\citeauthoryear{{Binney}}{{Binney}}{1978}]{1978MNRAS.183..501B}
{Binney} J.,  1978, \mn@doi [\mnras] {10.1093/mnras/183.3.501}, \href
  {https://ui.adsabs.harvard.edu/abs/1978MNRAS.183..501B} {183, 501}

\bibitem[\protect\citeauthoryear{{Binney}}{{Binney}}{2005}]{2005MNRAS.363..937B}
{Binney} J.,  2005, \mn@doi [\mnras] {10.1111/j.1365-2966.2005.09495.x}, \href
  {https://ui.adsabs.harvard.edu/abs/2005MNRAS.363..937B} {363, 937}

\bibitem[\protect\citeauthoryear{{Binney} \& {Mamon}}{{Binney} \&
  {Mamon}}{1982}]{1982MNRAS.200..361B}
{Binney} J.,  {Mamon} G.~A.,  1982, \mn@doi [\mnras] {10.1093/mnras/200.2.361},
  \href {https://ui.adsabs.harvard.edu/abs/1982MNRAS.200..361B} {200, 361}

\bibitem[\protect\citeauthoryear{{Binney} \& {Tremaine}}{{Binney} \&
  {Tremaine}}{2008}]{2008gady.book.....B}
{Binney} J.,  {Tremaine} S.,  2008, Galactic Dynamics: Second Edition.
Princeton University Press, Princeton, NJ, \url
  {https://books.google.co.uk/books?id=6mF4CKxlbLsC}

\bibitem[\protect\citeauthoryear{{Cappellari}}{{Cappellari}}{2002}]{2002MNRAS.333..400C}
{Cappellari} M.,  2002, \mn@doi [\mnras] {10.1046/j.1365-8711.2002.05412.x},
  \href {https://ui.adsabs.harvard.edu/abs/2002MNRAS.333..400C} {333, 400}

\bibitem[\protect\citeauthoryear{{Cappellari}}{{Cappellari}}{2008}]{2008MNRAS.390...71C}
{Cappellari} M.,  2008, \mn@doi [\mnras] {10.1111/j.1365-2966.2008.13754.x},
  \href {https://ui.adsabs.harvard.edu/abs/2008MNRAS.390...71C} {390, 71}

\bibitem[\protect\citeauthoryear{{Cappellari}}{{Cappellari}}{2016}]{2016ARA&A..54..597C}
{Cappellari} M.,  2016, \mn@doi [\araa] {10.1146/annurev-astro-082214-122432},
  \href {https://ui.adsabs.harvard.edu/abs/2016ARA&A..54..597C} {54, 597}

\bibitem[\protect\citeauthoryear{{Cappellari}}{{Cappellari}}{2020}]{Cappellari2020}
{Cappellari} M.,  2020, \mn@doi [\mnras] {10.1093/mnras/staa959}, \href
  {https://ui.adsabs.harvard.edu/abs/2020MNRAS.494.4819C} {494, 4819}

\bibitem[\protect\citeauthoryear{{Cappellari} \& {Copin}}{{Cappellari} \&
  {Copin}}{2003}]{2003MNRAS.342..345C}
{Cappellari} M.,  {Copin} Y.,  2003, \mn@doi [\mnras]
  {10.1046/j.1365-8711.2003.06541.x}, \href
  {https://ui.adsabs.harvard.edu/abs/2003MNRAS.342..345C} {342, 345}

\bibitem[\protect\citeauthoryear{{Cappellari} et~al.,}{{Cappellari}
  et~al.}{2007}]{2007MNRAS.379..418C}
{Cappellari} M.,  et~al., 2007, \mn@doi [\mnras]
  {10.1111/j.1365-2966.2007.11963.x}, \href
  {https://ui.adsabs.harvard.edu/abs/2007MNRAS.379..418C} {379, 418}

\bibitem[\protect\citeauthoryear{{Cappellari} et~al.,}{{Cappellari}
  et~al.}{2008}]{2008IAUS..245..215C}
{Cappellari} M.,  et~al., 2008, in {Bureau} M.,  {Athanassoula} E.,   {Barbuy}
  B.,  eds,  IAU Symposium Vol. 245, Formation and Evolution of Galaxy Bulges.
  pp 215--218 (\mn@eprint {arXiv} {0709.2861}),
  \mn@doi{10.1017/S1743921308017687}

\bibitem[\protect\citeauthoryear{{Cappellari} et~al.,}{{Cappellari}
  et~al.}{2013}]{2013MNRAS.432.1709C}
{Cappellari} M.,  et~al., 2013, \mn@doi [\mnras] {10.1093/mnras/stt562}, \href
  {https://ui.adsabs.harvard.edu/abs/2013MNRAS.432.1709C} {432, 1709}

\bibitem[\protect\citeauthoryear{{Davies}, {Efstathiou}, {Fall}, {Illingworth}
  \& {Schechter}}{{Davies} et~al.}{1983}]{1983ApJ...266...41D}
{Davies} R.~L.,  {Efstathiou} G.,  {Fall} S.~M.,  {Illingworth} G.,
  {Schechter} P.~L.,  1983, \mn@doi [\apj] {10.1086/160757}, \href
  {https://ui.adsabs.harvard.edu/abs/1983ApJ...266...41D} {266, 41}

\bibitem[\protect\citeauthoryear{{Emsellem}, {Monnet}  \& {Bacon}}{{Emsellem}
  et~al.}{1994}]{1994A&A...285..723E}
{Emsellem} E.,  {Monnet} G.,   {Bacon} R.,  1994, \aap, \href
  {https://ui.adsabs.harvard.edu/abs/1994A&A...285..723E} {285, 723}

\bibitem[\protect\citeauthoryear{{Emsellem} et~al.,}{{Emsellem}
  et~al.}{2007}]{2007MNRAS.379..401E}
{Emsellem} E.,  et~al., 2007, \mn@doi [\mnras]
  {10.1111/j.1365-2966.2007.11752.x}, \href
  {https://ui.adsabs.harvard.edu/abs/2007MNRAS.379..401E} {379, 401}

\bibitem[\protect\citeauthoryear{{Emsellem} et~al.,}{{Emsellem}
  et~al.}{2011}]{2011MNRAS.414..888E}
{Emsellem} E.,  et~al., 2011, \mn@doi [\mnras]
  {10.1111/j.1365-2966.2011.18496.x}, \href
  {https://ui.adsabs.harvard.edu/abs/2011MNRAS.414..888E} {414, 888}

\bibitem[\protect\citeauthoryear{{Gott}}{{Gott}}{1975}]{1975ApJ...201..296G}
{Gott} J.~Richard I.,  1975, \mn@doi [\apj] {10.1086/153887}, \href
  {https://ui.adsabs.harvard.edu/abs/1975ApJ...201..296G} {201, 296}

\bibitem[\protect\citeauthoryear{{Graham} et~al.,}{{Graham}
  et~al.}{2018}]{2018MNRAS.477.4711G}
{Graham} M.~T.,  et~al., 2018, \mn@doi [\mnras] {10.1093/mnras/sty504}, \href
  {https://ui.adsabs.harvard.edu/abs/2018MNRAS.477.4711G} {477, 4711}

\bibitem[\protect\citeauthoryear{{Illingworth}}{{Illingworth}}{1977}]{1977ApJ...218L..43I}
{Illingworth} G.,  1977, \mn@doi [\apjl] {10.1086/182572}, \href
  {https://ui.adsabs.harvard.edu/abs/1977ApJ...218L..43I} {218, L43}

\bibitem[\protect\citeauthoryear{{Jeans}}{{Jeans}}{1922}]{1922MNRAS..82..122J}
{Jeans} J.~H.,  1922, \mn@doi [\mnras] {10.1093/mnras/82.3.122}, \href
  {https://ui.adsabs.harvard.edu/abs/1922MNRAS..82..122J} {82, 122}

\bibitem[\protect\citeauthoryear{{Jenkins} \& {Binney}}{{Jenkins} \&
  {Binney}}{1990}]{1990MNRAS.245..305J}
{Jenkins} A.,  {Binney} J.,  1990, \mnras, \href
  {https://ui.adsabs.harvard.edu/abs/1990MNRAS.245..305J} {245, 305}

\bibitem[\protect\citeauthoryear{{Kormendy}}{{Kormendy}}{1982}]{1982ApJ...257...75K}
{Kormendy} J.,  1982, \mn@doi [\apj] {10.1086/159964}, \href
  {https://ui.adsabs.harvard.edu/abs/1982ApJ...257...75K} {257, 75}

\bibitem[\protect\citeauthoryear{{Kormendy} \& {Illingworth}}{{Kormendy} \&
  {Illingworth}}{1982}]{1982ApJ...256..460K}
{Kormendy} J.,  {Illingworth} G.,  1982, \mn@doi [\apj] {10.1086/159923}, \href
  {https://ui.adsabs.harvard.edu/abs/1982ApJ...256..460K} {256, 460}

\bibitem[\protect\citeauthoryear{{Krajnovi{\'c}} et~al.,}{{Krajnovi{\'c}}
  et~al.}{2018}]{2018MNRAS.477.3030K}
{Krajnovi{\'c}} D.,  et~al., 2018, \mn@doi [\mnras] {10.1093/mnras/sty778},
  \href {https://ui.adsabs.harvard.edu/abs/2018MNRAS.477.3030K} {477, 3030}

\bibitem[\protect\citeauthoryear{{Poci}, {Cappellari}  \& {McDermid}}{{Poci}
  et~al.}{2017}]{Poci2017}
{Poci} A.,  {Cappellari} M.,   {McDermid} R.~M.,  2017, \mn@doi [\mnras]
  {10.1093/mnras/stx101}, \href
  {https://ui.adsabs.harvard.edu/abs/2017MNRAS.467.1397P} {467, 1397}

\bibitem[\protect\citeauthoryear{{Rybicki}}{{Rybicki}}{1987}]{Rybicki1987}
{Rybicki} G.~B.,  1987, in {de Zeeuw} P.~T.,  ed.,  IAU Symposium Vol. 127,
  Structure and Dynamics of Elliptical Galaxies. D. Reidel, Dordrecht, p.~397,
  \mn@doi{10.1007/978-94-009-3971-4_41}

\bibitem[\protect\citeauthoryear{{Schwarzschild}}{{Schwarzschild}}{1979}]{Schwarzschild1979}
{Schwarzschild} M.,  1979, \mn@doi [\apj] {10.1086/157282}, \href
  {https://ui.adsabs.harvard.edu/abs/1979ApJ...232..236S} {232, 236}

\bibitem[\protect\citeauthoryear{{Scott} et~al.,}{{Scott}
  et~al.}{2013}]{2013MNRAS.432.1894S}
{Scott} N.,  et~al., 2013, \mn@doi [\mnras] {10.1093/mnras/sts422}, \href
  {https://ui.adsabs.harvard.edu/abs/2013MNRAS.432.1894S} {432, 1894}

\bibitem[\protect\citeauthoryear{{Shapiro}, {Gerssen}  \& {van der
  Marel}}{{Shapiro} et~al.}{2003}]{2003AJ....126.2707S}
{Shapiro} K.~L.,  {Gerssen} J.,   {van der Marel} R.~P.,  2003, \mn@doi [\aj]
  {10.1086/379306}, \href
  {https://ui.adsabs.harvard.edu/abs/2003AJ....126.2707S} {126, 2707}

\bibitem[\protect\citeauthoryear{{Spitzer} \& {Schwarzschild}}{{Spitzer} \&
  {Schwarzschild}}{1951}]{1951ApJ...114..385S}
{Spitzer} Lyman J.,  {Schwarzschild} M.,  1951, \mn@doi [\apj]
  {10.1086/145478}, \href
  {https://ui.adsabs.harvard.edu/abs/1951ApJ...114..385S} {114, 385}

\bibitem[\protect\citeauthoryear{{Thob} et~al.,}{{Thob}
  et~al.}{2019}]{2019MNRAS.485..972T}
{Thob} A. C.~R.,  et~al., 2019, \mn@doi [\mnras] {10.1093/mnras/stz448}, \href
  {https://ui.adsabs.harvard.edu/abs/2019MNRAS.485..972T} {485, 972}

\bibitem[\protect\citeauthoryear{{Thomas} et~al.,}{{Thomas}
  et~al.}{2009}]{2009MNRAS.393..641T}
{Thomas} J.,  et~al., 2009, \mn@doi [\mnras]
  {10.1111/j.1365-2966.2008.14238.x}, \href
  {https://ui.adsabs.harvard.edu/abs/2009MNRAS.393..641T} {393, 641}

\bibitem[\protect\citeauthoryear{{Toth} \& {Ostriker}}{{Toth} \&
  {Ostriker}}{1992}]{1992ApJ...389....5T}
{Toth} G.,  {Ostriker} J.~P.,  1992, \mn@doi [\apj] {10.1086/171185}, \href
  {https://ui.adsabs.harvard.edu/abs/1992ApJ...389....5T} {389, 5}

\bibitem[\protect\citeauthoryear{{Wang}, {Cappellari}, {Peng}  \&
  {Graham}}{{Wang} et~al.}{2020}]{2020MNRAS.495.1958W}
{Wang} B.,  {Cappellari} M.,  {Peng} Y.,   {Graham} M.,  2020, \mn@doi [\mnras]
  {10.1093/mnras/staa1325}, \href
  {https://ui.adsabs.harvard.edu/abs/2020MNRAS.495.1958W} {495, 1958}

\bibitem[\protect\citeauthoryear{{van de Sande} et~al.,}{{van de Sande}
  et~al.}{2017}]{vandeSande2017}
{van de Sande} J.,  et~al., 2017, \mn@doi [\apj] {10.3847/1538-4357/835/1/104},
  \href {https://ui.adsabs.harvard.edu/abs/2017ApJ...835..104V} {835, 104}

\makeatother
\end{thebibliography}


\bsp    
\label{lastpage}
\end{document}